\documentclass[a4paper,10pt]{amsart}
\usepackage{xcolor}
\usepackage[utf8]{inputenc}
\usepackage[a4paper, left=4.0cm, right=4.0cm, top=3.0cm, bottom=3.5cm, includefoot]{geometry}
\usepackage[
    hyperref, backend=biber, doi=false,url=false,sorting=none,backref=true]{biblatex}
\DefineBibliographyStrings{spanish}{andothers={\textsl{et al}\adddot}}
\renewbibmacro{in:}{}

\usepackage{amsmath, amssymb,stackrel,amsaddr}
\usepackage{hyperref}
\usepackage[english]{babel}
\usepackage{graphicx}

\hypersetup{colorlinks=true,linktoc=page,linkcolor=blue,citecolor=red,urlcolor=cyan}

\bibliography{bibliografia}

\title{The Snyder-de Sitter Scalar $\varphi^4_{\star}$ Quantum Field Theory in D=2}
\author{S.~A.~Franchino-Vi\~nas}
\address{Departamento de F\'isica, Facultad de Ciencias Exactas
Universidad Nacional de La Plata, C.C.\ 67 (1900), La Plata, Argentina.}

\address{Institut für Theoretische Physik, Universität Heidelberg, D-69120 Heidelberg, Germany.}

\email{safranchino@fisica.unlp.edu.ar}

\author{S. Mignemi}
\address{Dipartimento di Matematica e Informatica, Università di Cagliari, viale Merello 92, \\09123 Cagliari, Italy.}
\address{
INFN, Sezione di Cagliari, Cittadella Universitaria, 09042 Monserrato, Italy.}
\email{smignemi@unica.it}

\newcommand{\nocol}[1]{{#1}}

\begin{document}

\begin{abstract}
We study the two-dimensional version of a quartic self-interacting quantum scalar field on a curved and noncommutative space (Snyder-de Sitter).
We show that the model is renormalizable at the one-loop level and compute the beta functions of the related couplings.
The renormalization group flow is then studied numerically,
arriving at the conclusion that noncommutative-curved deformations can yield both relevant and irrelevant contributions to the one-loop effective action.
\end{abstract}
\maketitle

\section{Introduction}
Originally introduced by Snyder in his search for  a natural cutoff in Quantum Field Theories (QFT) \cite{Snyder:1946qz},
noncommutative geometry has become an important branch in mathematics \cite{Connes:1994yd}.
More important to us is the fact that noncommutative QFT has become one of the main tools in the attempt to
gather information and experience on possible effects of Quantum Gravity.
In few words, it materializes the conjecture that spacetime should show a granular behaviour at high enough energies \cite{Doplicher:1994tu}.
Since the literature on the field is really vast, we refer the interested reader to the reports \cite{Douglas:2001ba, Szabo:2001kg}
and references therein.

Recently, there has been an increasing interest in the consideration of noncommutative theories in curved spaces,
driven by astronomical observations, which indicate that we live in a universe that is not flat \cite{Aghanim:2018eyx}.
These proposals include a curved $\kappa$-Minkowski spacetime built from
Poisson-Lie algebras structures \cite{Gutierrez-Sagredo:2019ipf, Ballesteros:2019hbw},
de Sitter fuzzy space group-theoretically constructed \cite{Buric:2020wwk, Buric:2019yau, Buric:2018kdi},
deformations of QFTs in de Sitter using embeddings in higher-dimensional spaces \cite{Frob:2020ctp},
and Snyder curved spaces \cite{KowalskiGlikman:2004kp, Mignemi:2008fj, Mignemi:2015una}.

Following the last approach, in \cite{Franchino-Vinas:2019nqy} we have proposed an action for a scalar self-interacting QFT in Snyder-de Sitter space (SdS),
\nocol{employing techniques which were previously developed for QFT in Snyder space \cite{ Meljanac:2017ikx, Meljanac:2017grw, Meljanac:2017jyk, Vinas:2014exa}.
SdS was first introduced in \cite{KowalskiGlikman:2004kp} under the name of triply special relativity.
One major motivation for considering a QFT on it, is the fact that the modified kinetic term develops a harmonic term.
Such a term is well-known for its consequences in the renormalization properties of the Grosse-Wulkenhaar model \cite{Grosse:2019qps, Grosse:2005da, Grosse:2004by}},
curing the so-called UV/IR mixing problem \cite{Minwalla:1999px}.

In spite of this similarity, the renormalization flow in SdS is much more complicated.
At the one-loop level and at first order in the noncommutative and curvature parameters,
the effective action develops divergences and requires the introduction of new counterterms in the original action.
Even more relevant is the fact that the theory apparently has no fixed points.
{Instead, similarly to the Grosse-Wulkenhaar model and to the nonlocal, albeit covariant theory in \cite{Trinchero:2018gwe,Trinchero:2012zn}, it admits an asymptotically free regime.}

Nevertheless, the proposal remains interesting and deserves further study for three major reasons.
First, it provides a natural explanation for the harmonic term,
which was introduced by hand in the Grosse-Wulkenhaar model and could have a clear geometric/astronomical meaning,
given its link to the cosmological constant in our model.
Second, depending on the values of the coupling constants at low energies,
the curvature parameter could display a change of sign as a consequence of its renormalization flow.
Such a change of sign is compatible with observational data according to \cite{Akarsu:2019hmw,Akarsu:2020yqa}.
Third, it provides a way to circumvent both the Swampland conjecture as proposed in \cite{deAlwis:2019aud},
and (even if not mentioned in \cite{Franchino-Vinas:2019nqy}) the Gross-Coleman theorem,
which establishes that there is no scalar asymptotic free theory in $D=4$ if Poincaré invariance is taken for granted \cite{Coleman:1973sx}.

As a way to simplify the theoretical analysis of our model, of course at the expense of losing applicability,
we have decided to study a two-dimensional version of the quartic self-interacting scalar QFT in SdS.
It is our expectation that the two-dimensional model should preserve the main properties of the four-dimensional one,
as is the case for the Grosse-Wulkenhaar model.
In the present article we will show that, in the small-deformations  regime (meaning both small noncommutativity and curvature)
the model is one-loop renormalizable,
i.e.~no additional term should be added to those already present in the action.
Moreover, if one takes the computed one-loop beta functions at face value,
the presence of deformations allows some flows that differ from the usual commutative and flat ones.

The exposition of the article runs as follows.
In Sec.~\ref{sec:model} we rederive the main properties of the scalar self-interacting QFT in SdS in arbitrary dimensions.
Afterwards, in Sec.~\ref{sec:EA} we write down the divergent contributions and the beta functions for the two-dimensional case,
at the order of one-loop.
The study of the renormalization flow is explored in Sec.~\ref{sec:beta_functions}
and final remarks are made in Sec.~\ref{sec:conclusions}.
App.~\ref{app:coefficientsVW} is devoted to the expansion of the quartic potential for a small noncommutative parameter.

\section{A self-interacting scalar field in curved Snyder space}\label{sec:model}
In this section we will review the main features of the model of a self-interacting scalar field in SdS,
which was originally derived in \cite{Franchino-Vinas:2019nqy} and used in \cite{Franchino-Vinas:2019lyi}.
Let us recall that in the curved Snyder scenario in which we are interested,
the momentum and position operators do not sastify the canonical commutation relations, but close instead to a quadratic deformation of the algebra
\cite{KowalskiGlikman:2004kp},
 \begin{align}\label{eq:phase_space}
\begin{split}
 &[\hat{x}_i,\hat{x}_j]={\rm i}\beta^2 J_{ij},\qquad [\hat{p}_i,\hat{p}_j]={\rm i}\alpha^2 J_{ij},\\
 &[\hat{x}_i,\hat{p}_j]={\rm i}[\delta_{ij}+\alpha^2 \hat{x}_i\hat{x}_j+\beta^2\hat{p_j}\hat{p}_i+\alpha\beta (\hat{x}_j\hat{p}_i+\hat{p_i}\hat{x}_j)],
 \end{split}
\end{align}
where we have defined
\begin{align}\label{eq:lorentz_generator}
 J_{ij}=\frac{1}{2}(\hat x_i\hat p_j-\hat x_j\hat p_i+\hat p_j\hat x_i-\hat p_i\hat x_j).
\end{align}
At this point the reader may conjecture that the form of these commutators is rather arbitrary.
However, there are two significant symmetries behind them.

First of all, the operators $J_{ij}$ introduced in \eqref{eq:lorentz_generator} satisfy the algebra of the
Lorentz symmetries, even if the momentum and position operators are deformed.
Indeed, a direct computation shows that the commutators among them are those of the usual Lorentz algebra,
and their action on position and momentum operators is the one expected for vectors:
\begin{align}\label{eq:lorentz_action}
 \begin{split}
[J_{ij}, J_{kl}]&=i (\delta_{ik}J_{jl}-\delta_{il}J_{jk}-\delta_{jk}J_{il}+\delta_{jl}J_{ik}),\\
 [J_{ij}, \hat{p}_{k}]&=i (\delta_{ik} \hat{p}_{j}-\delta_{kj} \hat{p}_{i}),\\
 [J_{ij}, \hat{x}_{k}]&=i (\delta_{ik} \hat{x}_{j}-\delta_{kj} \hat{x}_{i}).
 \end{split}
 \end{align}

Second, there exists a duality symmetry under the exchange of position with momenta operators,
which can be written as
\begin{align}\label{eq:duality}
\alpha\hat x_i \leftrightarrow\beta\hat p_i.
\end{align}
This symmetry, originally proposed by Born as the reciprocity principle \cite{Born:1949}, has been more recently studied by Langman and Szabo \cite{Langmann:2002cc},
and exploited by Grosse and Wulkenhaar to obtain an all-order renormalizable and constructable noncommutative QFT \cite{Grosse:2005da}.

Turning back to \eqref{eq:phase_space}, one can readily notice that this algebra has two obvious limits: the limit $\alpha\to0$ corresponds to Snyder space (a noncommutative space or, equivalently, a curved momentum space), while $\beta\to0$ corresponds to de Sitter space (a curved space). By combining noncommutativity and curvature, we are led to deform also the commutation between positions and momenta as in \eqref{eq:phase_space}, in order to satisfy the Jacobi identities.

To construct our theory, we will define an action principle in SdS.
We will therefore consider, as customary, an action for the real scalar field,
\begin{align}\label{eq.S}
 S=S_K+S_I,
\end{align}
given by the sum of a kinetic term, $S_K$, and a  term of self-interaction, $S_I$,
which we describe in the following.

\subsection{The kinetic term}
Let us begin by considering the kinetic term.
A natural choice is to employ the square of the momentum operator  acting on the scalar;
in $D$ dimensions we write
\begin{align}\label{eq:action}
 S_K=\frac{1}{2}\int {\rm d}^Dx \,\varphi\, (\hat p^2+ m^2)\, \varphi.
\end{align}
The key point in the following is that, as can be guessed from \eqref{eq:phase_space}, $\hat p$ will not simply act as a derivative.

To elucidate this point, we will employ a fundamental property of SdS: by using a nonunitary, linear and evidently noncanonical transformation,
we can transform the algebra to that of Snyder space.
In \cite{Mignemi:2009zz,Mignemi:2015una} it has been shown that defining new position $X$ and momentum $P$ operators as
\begin{align}\label{eq:2snyder}
 \hat{x}_i=:X_i+t\, \frac{\beta}{\alpha} P_i, &\qquad \hat{p}_i=:(1-t)P_i -\frac{\alpha}{\beta} X_i,
\end{align}
one obtains the usual Snyder algebra in the form
\begin{equation}\label{eq:Snyder}
 [X_i,X_j]={\rm i}\beta^2 J_{ij}, \qquad [P_i,P_j]=0,\qquad [X_i,P_j]={\rm i}(\delta_{ij}+\beta^2P_iP_j),
\end{equation}
for any arbitrary value of the parameter $t$. There is a subtle point with this step:
as we will see later, the fact that this transformation is singular for small $\beta$ and $\alpha$
if $t$ is nonvanishing, prevents the commutative and flat limit.
Choosing $t=0$, we can conserve a smooth flat limit; the price that we have to pay is the fact that we won't be able to recover the curved commutative limit,
but only the commutative flat one.
An interesting question is whether these singularities are related to the existence of quantum phases,
i.e. related to singular transitions in the properties of the associated group, or are just a consequence of the chosen change of variables \cite{Nair:2000ii,Bellucci:2001xp}.

Another negative point is the fact that the Born principle does not hold for the variables $P$ and $X$, and the duality gets hidden.
Even if these two disadvantages are not minor ones,
working in Snyder space greatly simplifies the subsequent computations.
Given that \eqref{eq:2snyder} is the only linear transformation that accomplishes this job  without introducing further dimensional parameters
is \eqref{eq:2snyder}, we will use it.
Alternatively, as a consequence of the duality \eqref{eq:duality},
one could consider a transformation that goes from SdS to de Sitter space, corresponding to $t=1$.

Starting from \eqref{eq:2snyder}, we can define a non-linear realization for the operators in Snyder space, in terms of canonical operators $x$ and $p$
\begin{align}\label{eq:2canonical}
 P_i=:p_i=-{\rm i}\partial_i, \quad X_i=:x_i+\beta^2x_jp_jp_i=x_i-\beta^2x_j\partial_j\partial_i.
\end{align}
Since the operators $X_i$ defined in this way are non-hermitian, we need to apply a symmetrization as described in \cite{Meljanac:2017ikx},
\begin{align}\label{eq:symmetrization}
 X_i\rightarrow X_i=\hat x_i=x_i+\frac{\beta^2}{2} (x_jp_jp_i+p_ip_jx_j).
\end{align}
In this way we obtain a simple expression for the momentum:
\begin{align}\label{eq:phat}
 \hat{p}_i=p_i-\frac{\alpha}{\beta} x_i-\frac{\alpha \beta}{2} (x_jp_jp_i+p_ip_jx_j).
\end{align}
On physical grounds, we expect both $\alpha$ and $\beta$ to be small,
at least at small energies scales, since they are to be associated with the cosmological constant and the noncommutativity.
However, their quotient may be of order unity.
Taking this into account, we insert eq.~\eqref{eq:phat} into \eqref{eq:action};
up to order $\alpha^2$ and $\beta^2$ we get
\begin{align}\label{eq:kinetic}
 \begin{split}
S_K&\approx
 \frac{1}{2}\int {\rm d}^Dx \,\varphi\, \Bigg(p^2+\frac{\alpha^2}{\beta^2}x^2
 +2\alpha^2 x_jp_jp_ix_i+ m^2_{\text{eff}}\Bigg)\, \varphi,
 \end{split}
\end{align}
with an effective mass given by\footnote{Notice that in the published version of \cite{Franchino-Vinas:2019nqy} there are two mistakes: the effective mass appears in the action $S$ and not in its second variation, and the relevant factor is different by a factor five.
In spite of these modifications, the qualitative conclusions derived in \cite{Franchino-Vinas:2019nqy} are still valid.}
\begin{align}
 \label{eq:meff}
 m^2_{\text{eff}}:&=m^2-\frac{\alpha^2}{2}D(D+1).
\end{align}

Some remarks are now in order. At the end of this procedure,
the kinetic term has developed a harmonic contribution which is proportional to the quotient of the curvature and the noncommutativity.
This observation, which is similar to the one made in \cite{Buric:2009ss}, could be a crucial concept in the understanding of the Grosse-Wulkenhaar model \cite{Grosse:2005da}.
Indeed, such a term was originally introduced by hand in \cite{Grosse:2005da} in the so-called vulcanization of the $\lambda\phi^4_{\star}$ model in the Moyal plane, which has been proved to be all-order perturbatively renormalizable.

Additionally, the action \eqref{eq:kinetic} displays many common features that arise in commutative curved spaces.
First, it is not strange to encounter infrared divergences like the ones that could produce a negative mass as
in expression \eqref{eq:meff}.
In the commutative de Sitter space, one encounters similar problems arising from the representations of the isometry group $SO(D,1)$  \cite{Marolf:2012kh}; radiative problems for small masses are believed to be caused by a breaking of the perturbative expansion \cite{LopezNacir:2019ord}.
Moreover, terms involving both momentum and coordinates, such as the dilation-type operator $x\cdot p$,
arise directly from the Laplacian in curved space.

\subsection{The self-interaction term}
As interaction term, we choose a quartic potential for the scalar field $\varphi$:
\begin{align}\label{eq:interaction}
\begin{split}
 S_I&=\frac{\lambda}{4!}\int {\rm d}^Dx \; \varphi(\hat{x}) \big[ \varphi(\hat{x}) \big(\varphi^2(\hat{x})\big)\big].
 \end{split}
 \end{align}
As a consequence of the coordinate operators $\hat x$'s noncommutativity, this expression is not easy to handle.
In order to simplify the computations, we can make use of the noncommutative as well as nonassociative star product derived in \cite{Meljanac:2017ikx},
\begin{align}
 e^{{\rm i}k\cdot x}\star e^{{\rm i} q \cdot x}=\frac{e^{{\rm i}D(k, q)\cdot x}}{(1-\beta^2 k\cdot q)^{(D+1)/2}},
\end{align}
where the vector $D_{\mu}$ is given by
\begin{align}
 D_{\mu}(k,q):=\frac{1}{1-\beta^2 k \cdot q} \left[\left(1- \frac{\beta^2 k \cdot q}{1+\sqrt{1+\beta^2 k^2}}\right)k_{\mu}+ \sqrt{1+\beta^2 k^2} q_{\mu}\right].
\end{align}
Although the nonassociativity of the product means that the gathering of the different fields  in \eqref{eq:interaction} is not unique,
at  $\mathcal{O}(\beta^2)$ it plays no role.

We can now replace the position operators in \eqref{eq:interaction} by star products to finally obtain the equivalent expression
\begin{align}\label{eq:interaction2}
\begin{split}
 S_I
 &=\frac{\lambda}{4!}\int {\rm d}^Dx \; \varphi(x) \star \big[ \varphi(x) \star \big(\varphi(x) \star \varphi(x) \big)\big]
 \\
 &=\frac{\lambda}{4!} \int {\rm d}^Dx \,\left[\varphi^4 + \beta^2\, \varphi^4_{(1)}+\mathcal{O}(\beta^4)\right],
\end{split}
\end{align}
in terms of the first noncommutative correction
\begin{align}\label{eq.phi_fourth}
 \varphi^4_{(1)}:&= \frac{2}{3} \varphi^3    \Big( (D+2) +2 x^{\mu} \partial_{\mu} \Big) \partial^2\varphi.
\end{align}
Notice that this expansion involves the Laplacian as well as a dilation operator acting on $\varphi$.
\nocol{On the one side, the Laplacian is a semi-negative defined operator;
on the other side, the involved dilation operator is not Hermitian.
Actually, the operator $2x^\mu\partial_\mu+D$ is anti-Hermitian when acting on a complex field,
and could thus prompt a unitarity problem.
However, acting on a real field gives a real result.
This fact entails a possible loss of positivity in the Lagrangian once the noncommutative corrections become big enough
or, in other words, opens the door to possible instabilities and phases.
Apparently, this issue has passed unnoticed in the literature \cite{Franchino-Vinas:2018jcs, Franchino-Vinas:2019nqy, Meljanac:2017grw, Meljanac:2017jyk};
{instead, the positivity of the effective potential can be checked \cite{Meljanac:2017jyk}}.
We will not deal with it in the present article, leaving it for a future presentation.
}

\section{One-loop contribution to the effective action}\label{sec:EA}
Our focus will be centered on the computation of the divergent one-loop contributions arising for the SdS model, cf.~eq.~\eqref{eq:action},
from which the running of the coupling constants in the \nocol{Modified Minimal Subtraction} scheme can be read.
One economic way to do so is by employing the Worldline Formalism, which is closely related to heat-kernel techniques.
The noncommutative version of this technique \cite{Bonezzi:2012vr} has been extended to consider all-order harmonic contributions in the study of the Grosse-Wulkenhaar model \cite{Vinas:2014exa};
it has also proven helpful even in presence of terms terms with higher momentum powers in the action \cite{Franchino-Vinas:2018jcs}.

Basically, the computation goes at follows. The one-loop contribution to the effective action can be written in terms of the classical field $\phi$ as
\begin{align}
\Gamma_{1-\text{loop}}[\phi] = S[\phi]-\frac{1}{2} \int_{0}^{\infty}\frac{{\rm d}T }{T}\text{Tr} \left( e^{-T \,\delta^2 S}\right).
\end{align}
In other words, it is given by the heat-kernel of the operator $\delta^2 S$, the second variation of the action.
For a quartic potential in SdS, a master equation has been obtained in \cite{Franchino-Vinas:2019nqy}, considering a first-order expansion in the noncommutative parameter $\beta$ and the curvature $\alpha$. Let us recall that the Weyl-ordered expression for the second variation of $S$, which is needed in the Worldline Formalism  \cite{Bastianelli:1992ct} and is denoted by a $W$ subscript, is given by
\begin{align}\label{eq:variation_SW}
 \delta^2 S_{W}=p^2+\omega^2x^2
 +\alpha^2 (x_ix_jp_jp_i)_S+m^2+V_W.
\end{align}
In this expression the  Weyl-ordered potential $V_W$ reads
\begin{align}\label{eq:VW}
\begin{split}
 V_{W}:&=-\frac{1}{2}\frac{\lambda}{4!}\int  \frac{{\rm d}^D q_1 {\rm d}^Dq_2}{(2\pi)^{2D}} e^{ {\rm i} x ( q_1+q_2)} \left[4!\,  +\beta^2 \left(\alpha_{\mu\nu}' p^{\mu} p^{\nu}+\beta_{\mu}' p^{\mu}+\gamma'\right)\right] \tilde{\phi}_{1}\tilde{\phi_2},\\
 \end{split}
\end{align}
in terms of $\tilde \phi_i:=\tilde\phi(q_i)$, the Fourier transform of the classical field, and the tensorial coefficients $\alpha_{\mu\nu}$, $\beta_{\mu}$ and $\gamma$, which are functions\footnote{These tensorial coefficients, $\alpha_{\mu\nu}$ and $\beta_{\mu}$, should not be confused with the curvature and noncommutative parameters, $\alpha$ and $\beta$.} of the momenta $q_1,\,q_2$ and derivatives acting on $\tilde\phi_i$, reported in  App.~\ref{app:coefficientsVW}.
The potential term, even if it contains derivatives of the classical field $\phi$, is local at the order in  which we are working.

Now we can work in dimensional regularization defining $D=:2-\varepsilon$, and retain only the divergent contributions for $\varepsilon\rightarrow 0$.
A direct computation of the expansion gives
\begin{align}
\begin{split}
\Gamma_{1-\text{loop}}^{\text{div}}=\frac{\lambda}{48\pi \varepsilon }&\int {\rm d}^2x \left [
\frac{6 \alpha ^2 m^2}{\omega ^2}+8 \beta ^2 m^2+6+x^2 \left(9 \alpha ^2+16 \beta ^2 \omega ^2\right)
\right] \phi^2(x)
\\
&\hspace{3cm}+\frac{\lambda^2}{96\pi\varepsilon}\int {\rm d}^2x \left[ \frac{3 \alpha ^2}{\omega ^2}+4 \beta ^2 \right] \phi^4(x),
\end{split}
\end{align}
telling us that in order to have a well-defined theory we should renormalize just the mass $m$, the frequency $\omega$ and the coupling constant $\lambda$
by introducing appropiate counterterms.

In particular, this means that in two dimensions the theory is renormalizable  at the one-loop level.
\nocol{This result is non-trivial, since a power counting argument would say that the theory is non-renormalizable
or, analogously, that some operators are classically relevant in the UV}.
This is in contrast with the four-dimensional case \cite{Franchino-Vinas:2019nqy},
where we had to perform a renormalization of the noncommutative parameter and of the curvature,
and even to introduce some additional terms in the original action.

It is now straightforward to introduce the necessary counterterms and compute the beta functions
for the dimensionful couplings.
Introducing the renormalization scale $\mu$ and defining $\beta_{x}=\frac{\partial x}{\partial \log \mu}$,
we are lead to\footnote{\nocol{We are still working at $\mathcal{O}(\alpha^2,\,\beta^2)$.} }
\begin{align}
\begin{split}\label{eq:betafunctions}
 \beta_{\omega^2}
 &=-\frac{\lambda}{12\pi } \left(9 \alpha ^2+8 \beta ^2 \omega ^2\right),\\
 \beta_{ m^2_{\text{eff}}}
 &=-\frac{\lambda}{12\pi }  \left( \frac{6 \alpha ^2  m^2}{\omega ^2}+4 \beta ^2  m^2+6\right),\\
 \beta_{\lambda}
 &=- \frac{\lambda^2  }{\pi}\left[ \frac{3 \alpha ^2}{\omega ^2}+2 \beta ^2 \right],
 \\
 \beta_{\beta^2}
 &=0,
 \\
 \beta_{\alpha^2}
 &= 0.
 \end{split}
\end{align}
In any case remember that, when we consider the Callan-Symanzik equation \cite{peshkin}, we need to add a contribution proportional to the classical dimension
of the coupling in every single differential equation in \eqref{eq:betafunctions}.

 \section{Analysis of the beta functions}\label{sec:beta_functions}
First of all, let us look for fixed points (FPs) of the system at the one-loop level, which are obtained by equating \nocol{
to zero the system of eqs.~\eqref{eq:betafunctions} with the addition of the classical dimensions.}
 Since $\alpha$ and $\beta$ have no contribution from anomalous dimensions to compensate their classical dimensions, the only solution for them is the trivial one.
 Replacing these values in the remaining equations, we get just one solution, namely the trivial Gaussian FP.
There is however a subtlety: as a consequence of our exact computation in $\omega$, the beta functions contain inverse powers of the frequency.
 This means that the FP can only be obtained dinamically, i.e.~we can not just set our parameters to the values of the FP, since at that point the expressions are not well-defined.
 Hence, our one-loop computation should be trusted only if the evolution is given in a certain region of the flow in the $(\alpha,\omega)$ plane,
 where the quotient $\frac{\alpha}{\omega}$ remains bounded.

 In order to study the system of coupled differential equations, we will proceed numerically in order to tackle the nontrivial ones.
 In the following the vector $v_i=(\omega^2_i,m_i^2,\alpha^2_i,\beta_i^2,\lambda_i)$ will label the initial conditions at a scale $\mu=1$ in arbitrary units.

 First of all, let us recall the commutative and flat result in $D=2$ without harmonic term,
 remembering that one should take the $\alpha\rightarrow 0$ limit before the $\omega\rightarrow 0$.
 In this case, the only nontrivial running corresponds to the mass.
 It is easy to see that the Gaussian FP is actually an attractor in the UV,
 so that the theory is asymptotically safe.
 Instead, if one trusts these one-loop results, the theory would be strongly coupled in the IR.
 The addition of a harmonic term introduces no new effect.

\begin{figure}[h!]
 \begin{minipage}{0.48\textwidth}

 \begin{center}
 \includegraphics[width=1.05\textwidth]{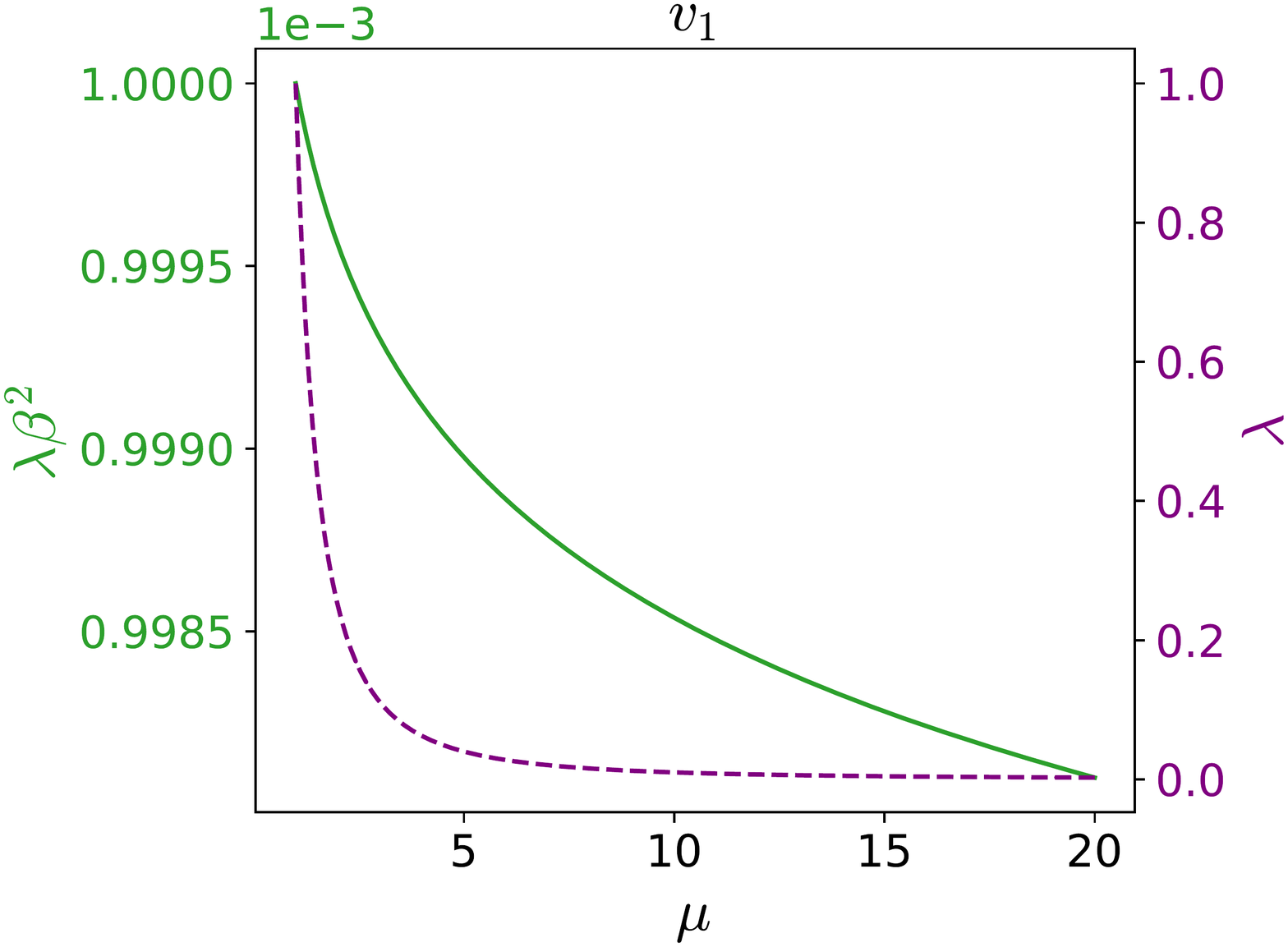}
 \end{center}
\end{minipage}
 \begin{minipage}{0.48\textwidth}

 \begin{center}
 \includegraphics[width=1.05\textwidth]{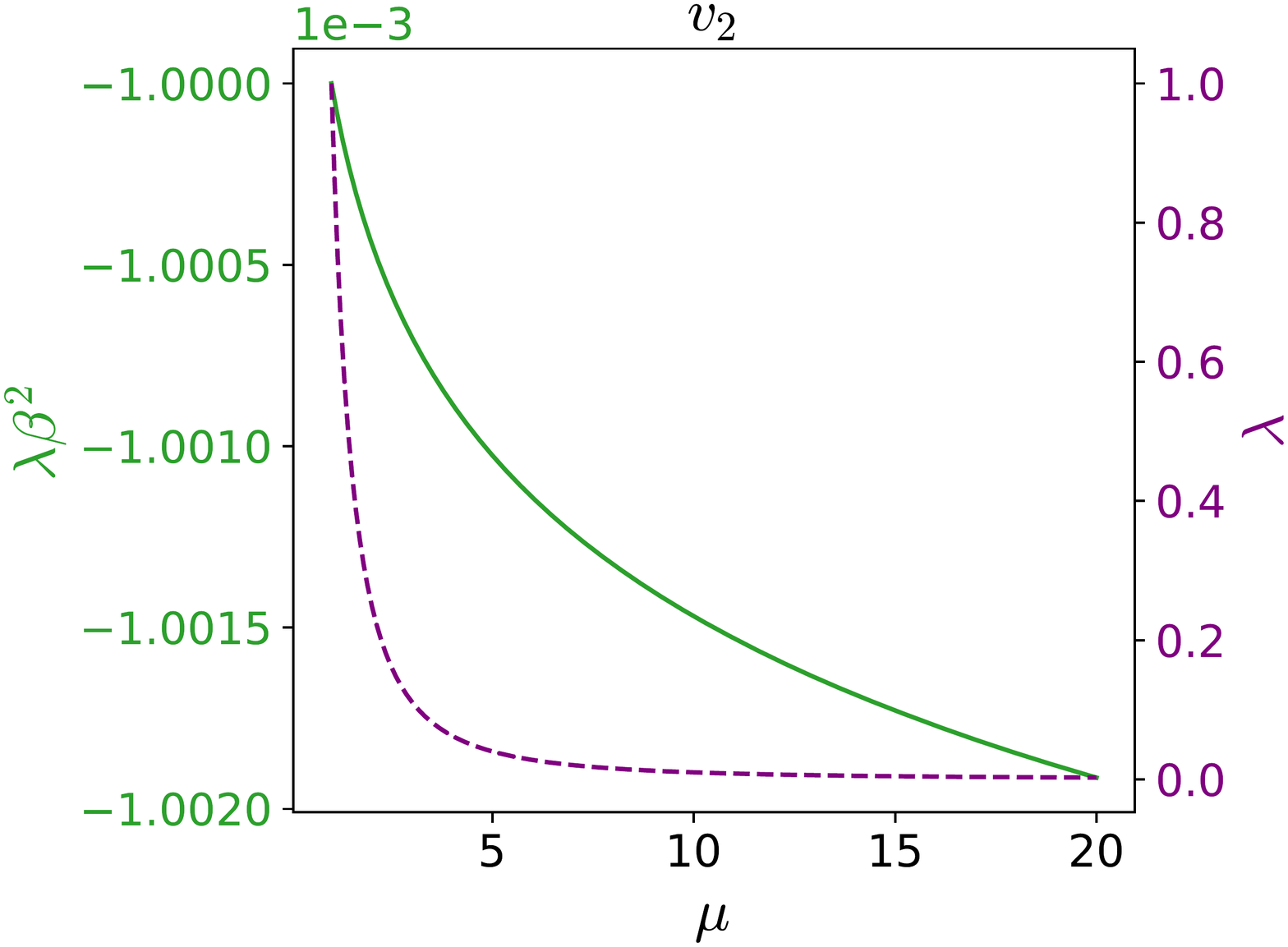}
 \end{center}
\end{minipage}

 \caption{The product $\lambda\beta^2$ (green continuous line) and the coupling constant $\lambda$ (purple dashed line) as a function of the energy $\mu$ in arbitrary units.
 The plot on the left corresponds to the choice of parameters $v_1=(0.5,\,1,\,0,\,10^{-3},\,1.)$, while the plot on the right belongs to $v_2=(0.5,\,1,\,0,\,-10^{-3},\,1.)$.
 }
 \label{fig:running_noncommutative}
\end{figure}

 On the other side, the introduction of the noncommutativity involves an operator which at first glance,
 given the trivial scaling of $\beta$,
 is relevant in the UV.
 However, the coupling involved in the noncommutative contribution to the potential is $\lambda\beta^2$;
 thus, there exists a dispute between the respectively decreasing and increasing behaviour of $\lambda$ and $\beta$.
 One can numerically see that the product in which we are interested tends to zero for large energies.
 As an example, choosing $v_1:=(0.5,\,1,\,0,\,10^{-3},\,1.)$, we have plotted both $\lambda\beta^2$ (in green solid line) and $\lambda$ (in purple dashed line)
 in the left panel of Fig.~\ref{fig:running_noncommutative}.
 Consequently the theory is asymptotically free, even if the noncommutative contribution becomes more important.

 The situation becomes unstable if instead of considering Snyder space we analyze a change of sign in $\beta^2$,
 i.e. we  consider anti-Snyder space. This happens  even in the case of a small noncommutativity.
 We depict this setting in the right panel of Fig.~\ref{fig:running_noncommutative}, with the choice $v_2=(0.5,\,1,\,0,\,-10^{-3},\,1.)$.
 Under these circumstances the anomalous dimension of $\lambda$, even if  not big enough to
 counteract the classical dimension, generates a growth in the absolute value of $\lambda\beta^2$.
 In other words, the operator associated with the noncommutative sector of the interaction will become relevant in the UV
 if we can extrapolate this one-loop computation.

\begin{figure}[h!]
  \begin{minipage}{0.48\textwidth}
 \begin{center}
 \includegraphics[width=1.05\textwidth]{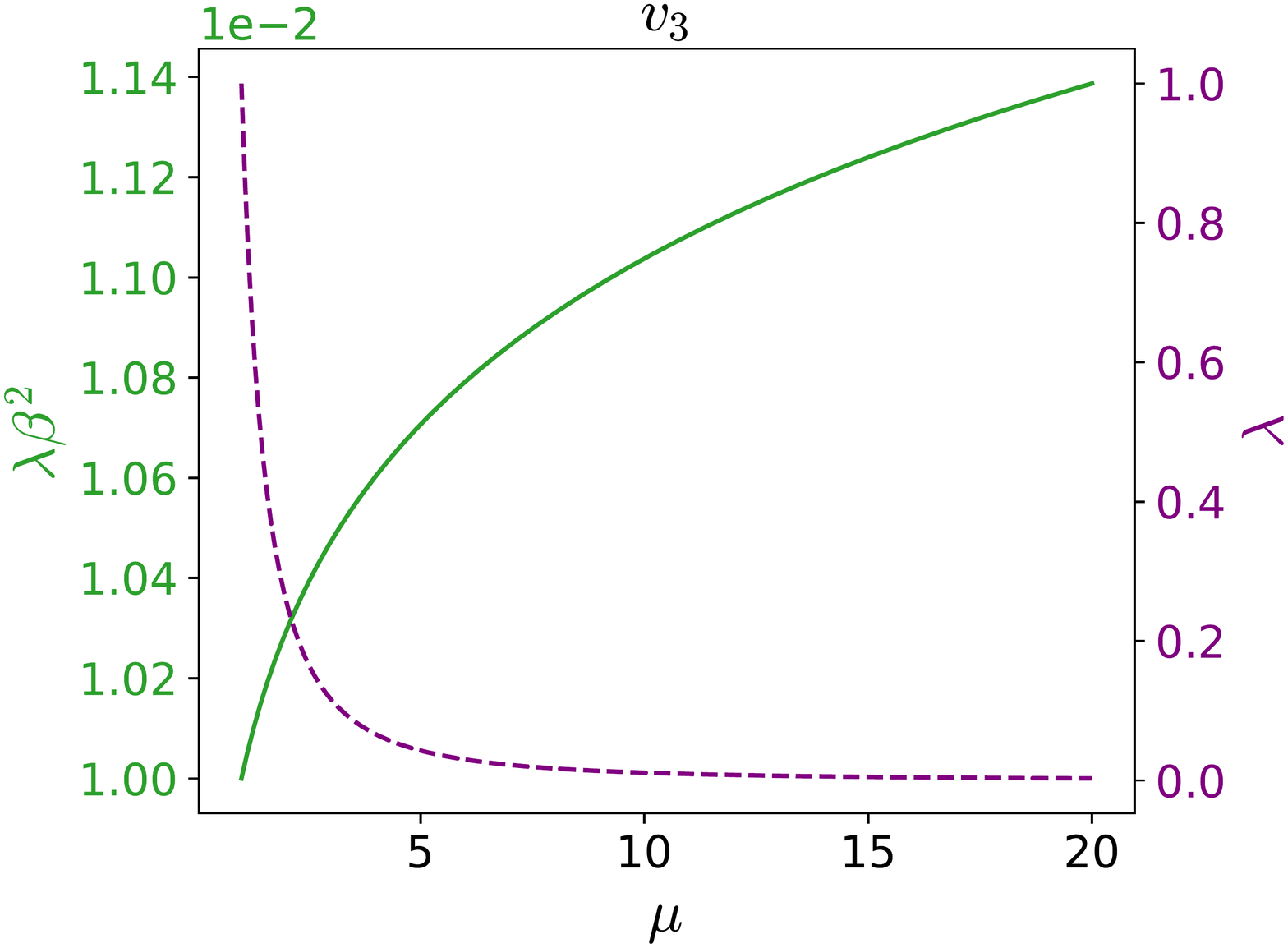}
 \end{center}

 \end{minipage}
 \caption{The product $\lambda\beta^2$ (green continuous line) and the coupling constant $\lambda$ (purple dashed line) as a function of the energy $\mu$ in arbitrary units,
 corresponding to the parameters $v_3=(0.1,\,1,\,-0.005,\,0.01,\,1)$.}
 \label{fig:running_oppositesign}

 \end{figure}

Let us now come to the full noncommutative and curved theory.
Although the consideration of a curvature $\alpha$ of the same
positive sign\footnote{Recall that $\alpha^2>0$ should be associated with a de Sitter geometry,
while $\alpha^2<0$ corresponds to an anti-de Sitter one.}
as $\beta$ leads always to a vanishing coupling $\lambda$ at high energies,
once we consider a negative curvature or noncommutativity the situation is not clearly defined.
This can be already seen from eq. \eqref{eq:betafunctions}, where the sign of the derivative
depends on the relative magnitude of $\alpha$ and $\beta$,
with an effect of the curvature parameter that could be enhanced by the frequency.
To be explicit, consider the Snyder-anti de Sitter case, where the initial values of $\alpha$ and $\beta$ are not so small,
viz.~$v_3:=(0.1,\,1,\,-0.005,\,0.01,\,1)$.
This case yields the situation depicted in Fig.~\ref{fig:running_oppositesign},
with an asymptotically free commutative contribution to the potential,
and a noncommutative contribution that is relevant in the UV if the one-loop is trustable enough.

Finally, let us mention an interesting alternative possibility related to the study of the beta functions,
cf.~\eqref{eq:betafunctions}.
\nocol{Following the results in \cite{Grosse:2005da} we could define dimensionless couplings that scale
either with the curvature or the noncommutative parameter.
This is natural, at least at the one-loop level, since they have no anomalous dimension.
This will not imply a change in the results obtained; it just renders the comparison with \cite{Grosse:2005da, Grosse:2004by} simpler.}

Perhaps the most natural rescaling would be $\omega\rightarrow \frac{\alpha}{\beta}\tilde\omega$, rendering the frequency parameter dimensionless.
Afterwards, some of the resulting beta functions could be made to vanish for some values of the parameters,
if the relative sign of $\omega$ with respect to $\alpha$ and $\beta$ is chosen appropiately.
In any case, this will allow a vanishing beta function for the mass and for either the frequency or the coupling (but never for both $\omega$ and $\lambda$).
Consequently, the fixed point structure in \cite{Grosse:2004by} will not be totally reproduced.
\nocol{Indeed, in \cite{Grosse:2004by} both the beta functions for the coupling and the dimensionless frequency vanish at the dual point,
while the beta function for the mass remains non-null.}

 \section{Conclusions}\label{sec:conclusions}

 We have shown that in two dimensions the SdS QFT is renormalizable at the one-loop level,
 contrary to what happens in the four-dimensional case.
 Indeed, we have seen that one needs to renormalize only the frequency, the mass and the coupling constant,
without the introduction of new terms in the action.

 Additionally, we have computed the beta functions of all the couplings involved.
 The frequency, the mass, the coupling constant $\lambda$ and $\alpha$ generically tend to zero as the energy increases.
 Instead, the operator corresponding to the noncommutative deformation of the potential,
 i.e.~the term  proportional to $\beta^2$,
 can be either relevant or irrelevant depending on the initial conditions.

 The analysis also shows that the system possesses no fixed points.
\nocol{Although one can introduce by hand the relative scale of one of the parameters,
 for example by replacing $\omega\rightarrow \frac{\alpha}{\beta}\tilde\omega$,
 and then study the running of the dimensionless parameter $\tilde\omega$,
 one would need to introduce also an additional relative sign in order to allow the vanishing of some beta functions.
 In any case, the pattern of fixed point of the Grosse-Wulkenhaar \cite{Grosse:2004by} is never attained.
 }

 \nocol{
 One possibility that has not yet been studied in SdS is a possible reformulation in terms of a matrix model,
 which was one of the keys of the success in \cite{Grosse:2005da}.
 This could render the renormalization properties more explicit.
}

\nocol{ Another interesting idea could be to consider an associative realization of the model, following \cite{Meljanac:2020fde}.
This would however require an additional compatification of the extra dimensions.}

 More generally, our model is intended to be an effective field theory in a noncommutative and curved space.
 One of the main features of any \nocol{Effective Field Theory} regards the identification of its underlying symmetries,
 in order to include in the action every possible term compatible with it.
 In our case, more focus should be made in the future on Born duality \cite{Born:1949},
 which is patent in our starting algebra \eqref{eq:phase_space}
 and lost in the building of the theory.
 Although, as can be seen from \eqref{eq:kinetic},
 the kinetic term in the SdS action develops several terms that treat $p$ and $x$ on equal footing,
 the potential term does not satisfy such a symmetry, at least in this small-deformations expansion.
 We believe that this is an important point on which one should focus, since it is a related symmetry, Langmann-Szabo's symmetry \cite{Langmann:2002cc},
 that is behind the properties of the celebrated Grosse-Wulkenhaar model \cite{Grosse:2005da}.
 To achieve this end, one should presumably figure out a method that does not rely on the transformations \eqref{eq:2snyder}.
 Such issue is currently investigated.

\section{Acknowledgements}
The authors thanks H. Grosse for his useful comments regarding stability and Hermiticity problems.
S.A.F. is grateful to G. Gori and the Institut für Theoretische Physik, Heidelberg, for their kind hospitality.
SAF acknowledges support from Project 11/X748 and Subsidio a Jóvenes Investigadores 2019, UNLP.
The authors would like to acknowledge networking support by the COST Action CA18108.

\appendix
\section{Coefficients in the expansion of the potential $V_W$}\label{app:coefficientsVW}
The tensorial coefficients $\alpha'_{\mu\nu}$, $\beta'_{\mu}$ and $\gamma'$, which are used in eq. \eqref{eq:VW},
have been derived for the first time in \cite{Franchino-Vinas:2018jcs}.
As a matter of completion, we write them explicitly, omitting the factors of the Fourier-transformed fields $\tilde\phi_i$,
which were taken out as overall factors in \eqref{eq:VW}:
\begin{align}
\begin{split}
 \alpha'_{\mu\nu}&=-8 (s_1+s_2) \left(2 (q_1+q_2)^{\nu}\partial_{q_1^{\mu}}+ (q_1+q_2)\cdot \partial_{q_1} \delta_{\mu\nu}+ (D+2)\delta_{\mu\nu}\right),\\
 \beta'_{\mu}&=0\,,\\
 \gamma'&=-2(s_1+s_2)  \Big[ 4(2q_1\cdot q_2+q_2^2)(q_1\cdot \partial_{q_1})+4(2q_1\cdot q_2+q_1^2)(q_2\cdot \partial_{q_1})\\
 &\hspace{2cm} -3 (q_1+q_2)^2 (q_1+q_2)\cdot \partial_{q_1}-  (2 + D) (q_1^2-2 q_1 q_2 - 3  q_2^2 ) \Big].
\end{split}
\end{align}

In these expressions, the derivatives are intended to apply solely to the right,
what in our case means only on the $\tilde\phi$ factors.
\printbibliography

\end{document}